\documentclass[runningheads]{svmult}

\usepackage{makeidx}   
\usepackage{graphicx}  
\usepackage{subeqnar}  
\usepackage{multicol}  
\usepackage{physprbb}  
\makeindex             

\newcommand{\greeksym}[1]{{\usefont{U}{psy}{m}{n}#1}}
\newcommand{\uDkorn}{\mbox{\greeksym{D}}}

\begin{document}
\title*{Pinning Down Gravitational Settling}
\toctitle{Pinning Down Gravitational Settling}
%
%
\titlerunning{Pinning Down Gravitational Settling}
%
\author{Andreas J. Korn\inst{1}
\and Nikolai Piskunov\inst{1}
\and Frank Grundahl\inst{2}
\and Paul Barklem\inst{1}
\and Bengt Gustafsson\inst{1}}
\authorrunning{A.J. Korn et al.}
%
%
\institute{Uppsala Astronomical Observatory, University of Uppsala, Uppsala, Sweden
\and Department of Astronomy, University of \AA rhus, \AA rhus, Denmark}

\maketitle              

\begin{abstract}
We analyse high-resolution archival UVES data of turnoff and subgiant stars in the nearby globular cluster NGC\,6397 ([Fe/H]\,$\approx$\,$-$2). Balmer-profile analyses are performed to derive reddening-free effective temperatures. Due to the limited S/N and uncertainties related to blaze removal, we find the data quality insufficient to exclude the existence of gravitational settling. If the newly derived effective temperatures are taken as a basis for an abundance analysis, the photospheric iron (Fe\,{\sc ii}) abundance in the turnoff stars is 0.11\,dex lower than in the (well-mixed) subgiants.
\end{abstract}

Practically all sophisticated stellar evolution models predict the existence of processes altering photospheric abundances on long timescales (see e.g. Pinsonneault, these proceedings). For example, Richard et al. \cite{richard} predict iron abundances in turnoff stars of NGC\,6397 to be lower by 0.2\,dex than in red giants.

\begin{figure}[!h]
\vspace*{-0.2cm}
\begin{center}
\hspace*{0.2cm}
\includegraphics[width=.68\textwidth,angle=90]{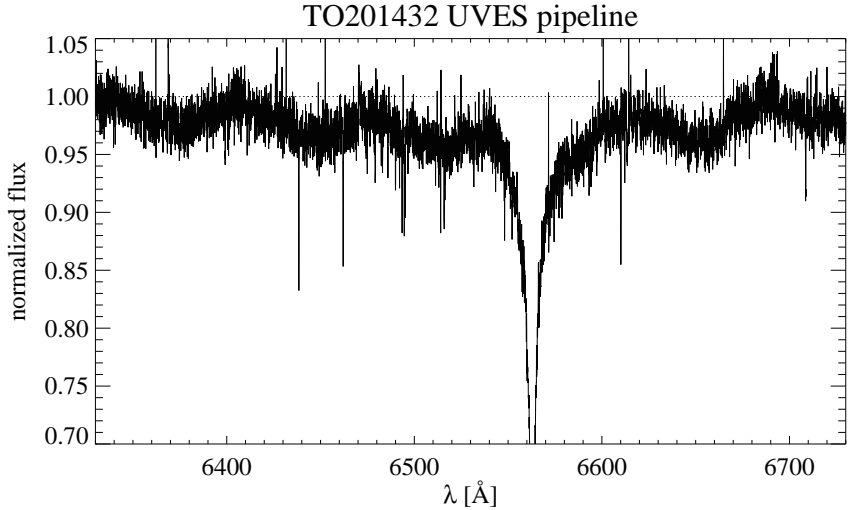}
\end{center}
\vspace*{-1.4cm}
\caption[]{Co-addition of four UVES pipeline spectra of NGC\,6397/TO201432 (observing dates 2000-06-18 and 22, two spectra per night). The resulting spectrum was arbitrarily normalized at 6410 and 6690\,\AA. As blaze residuals are not properly accounted for in the pipeline order merging, the \'{e}chelle order pattern is clearly visible in the merged spectrum. With an amplitude of 2\,\%, these instrumental artifacts do not allow to derive Balmer-profile temperatures to better than 200--300\,K.}
\label{pipeline}
\end{figure}


\begin{figure}[!h]
\vspace*{-0.2cm}
\begin{center}
\hspace*{0.2cm}
\vspace*{-1.4cm}
\includegraphics[width=.68\textwidth,angle=90]{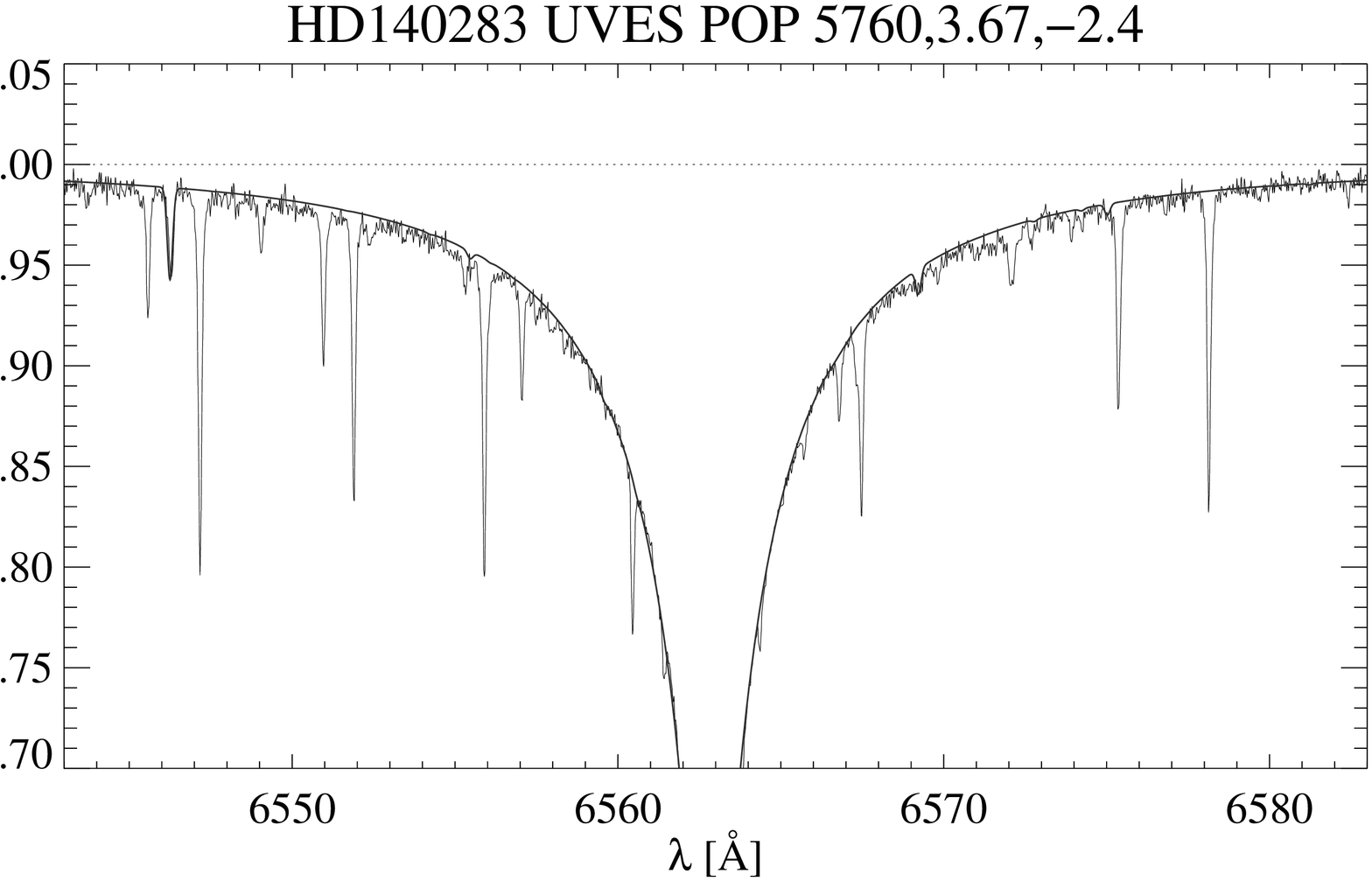}
\includegraphics[width=.68\textwidth,angle=90]{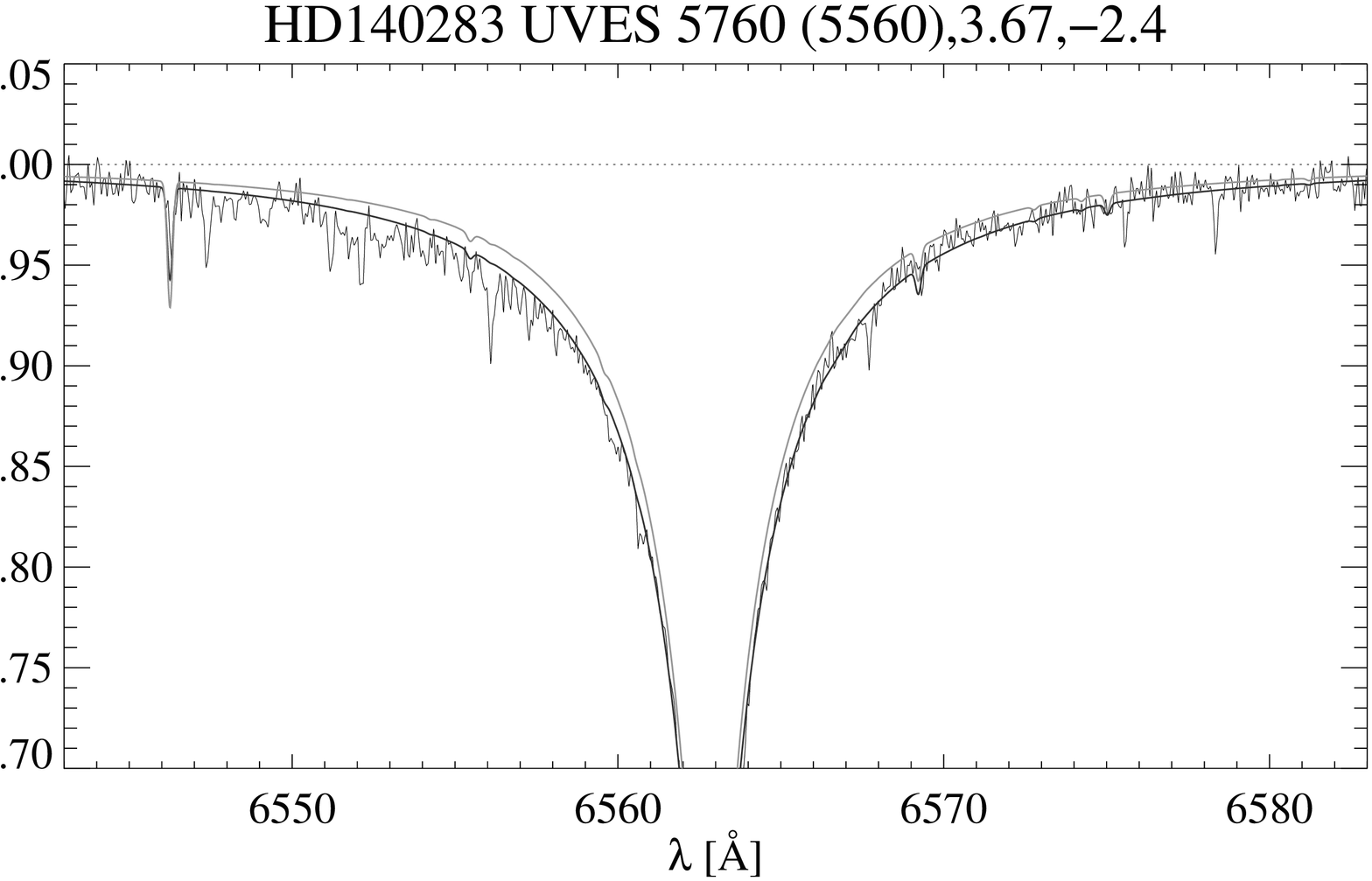}
\end{center}
\vspace*{-1.2cm}
\caption[]{Top panel: Observed (UVES POP \cite{UVESpop}) and synthetic H$\alpha$ profiles of the field star HD\,140283. The unmerged data (2001-07-09, frame 570) was retrieved from \cite{UVESpop} and rectified using a parabola interpolated from the continua of adjacent orders. A best-fit $T_{\rm eff}$ of 5760\,K (at log\,$g$\,=\,3.67 and [Fe/H]\,=\,$-$2.4, neither critical) is indicated.\\
Bottom panel: Observed and synthetic H$\alpha$ profiles for HD\,140283 (black: 5760\,K, grey: 5560\,K). The data was retrieved from the VLT archive (observing date 2000-06-15), reduced using REDUCE and rectified using parabolic fits to the continua in adjacent orders. Notice the variable telluric features which somewhat suppress the blue wing. An effective temperature of 5560\,K (best estimate of \cite{gratton}) is clearly too low, 5760\,K (derived from the UVES POP spectrum) too high. At a S/N of 150, observational systematic errors are thus of the order of at least 100\,K.}
\label{hd140283}
\end{figure}


\begin{figure}[!h]
\vspace*{-0.2cm}
\begin{center}
\hspace*{0.2cm}
\vspace*{-1.4cm}
\includegraphics[width=.68\textwidth,angle=90,clip=]{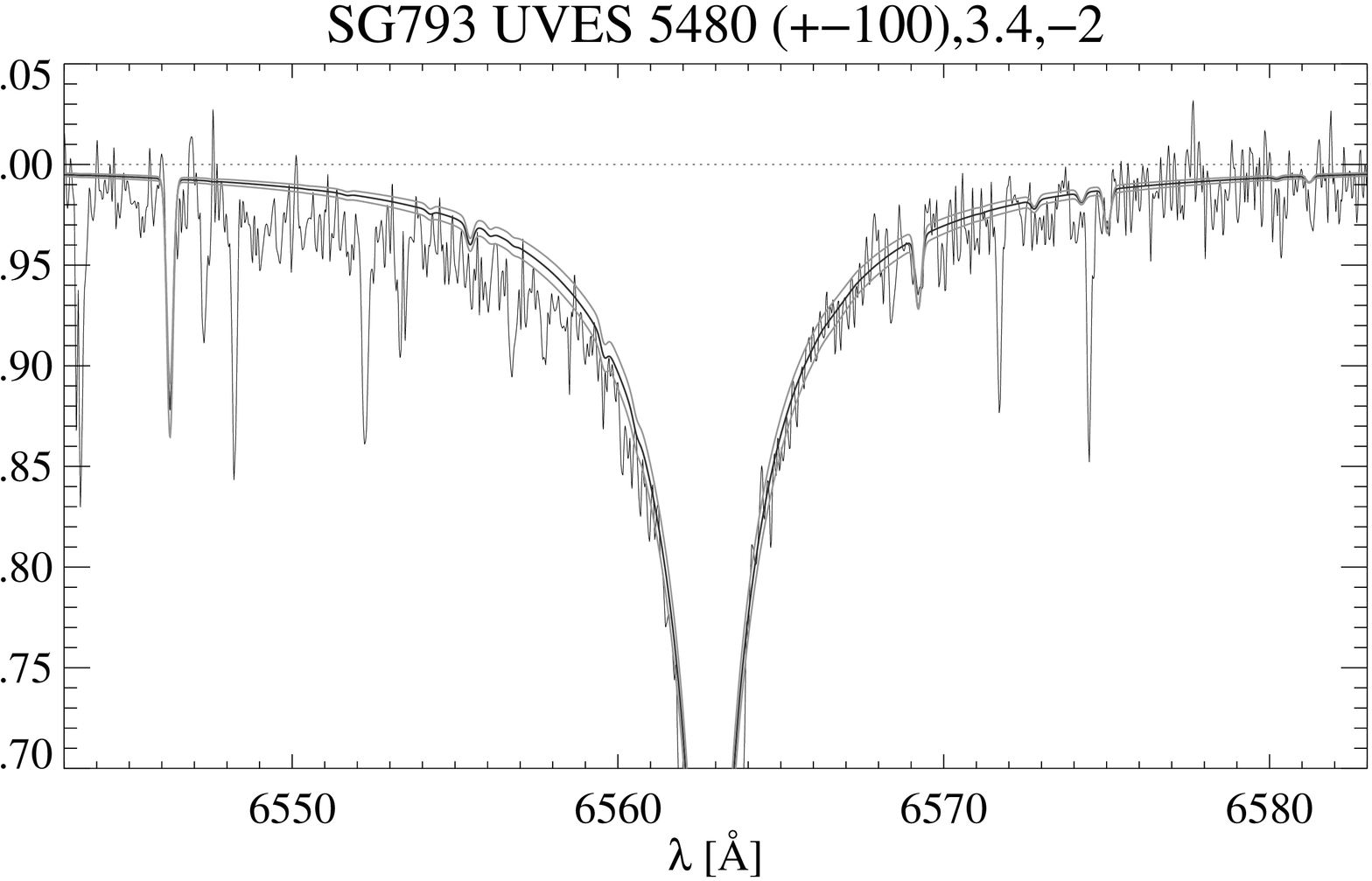}
\includegraphics[width=.68\textwidth,angle=90,clip=]{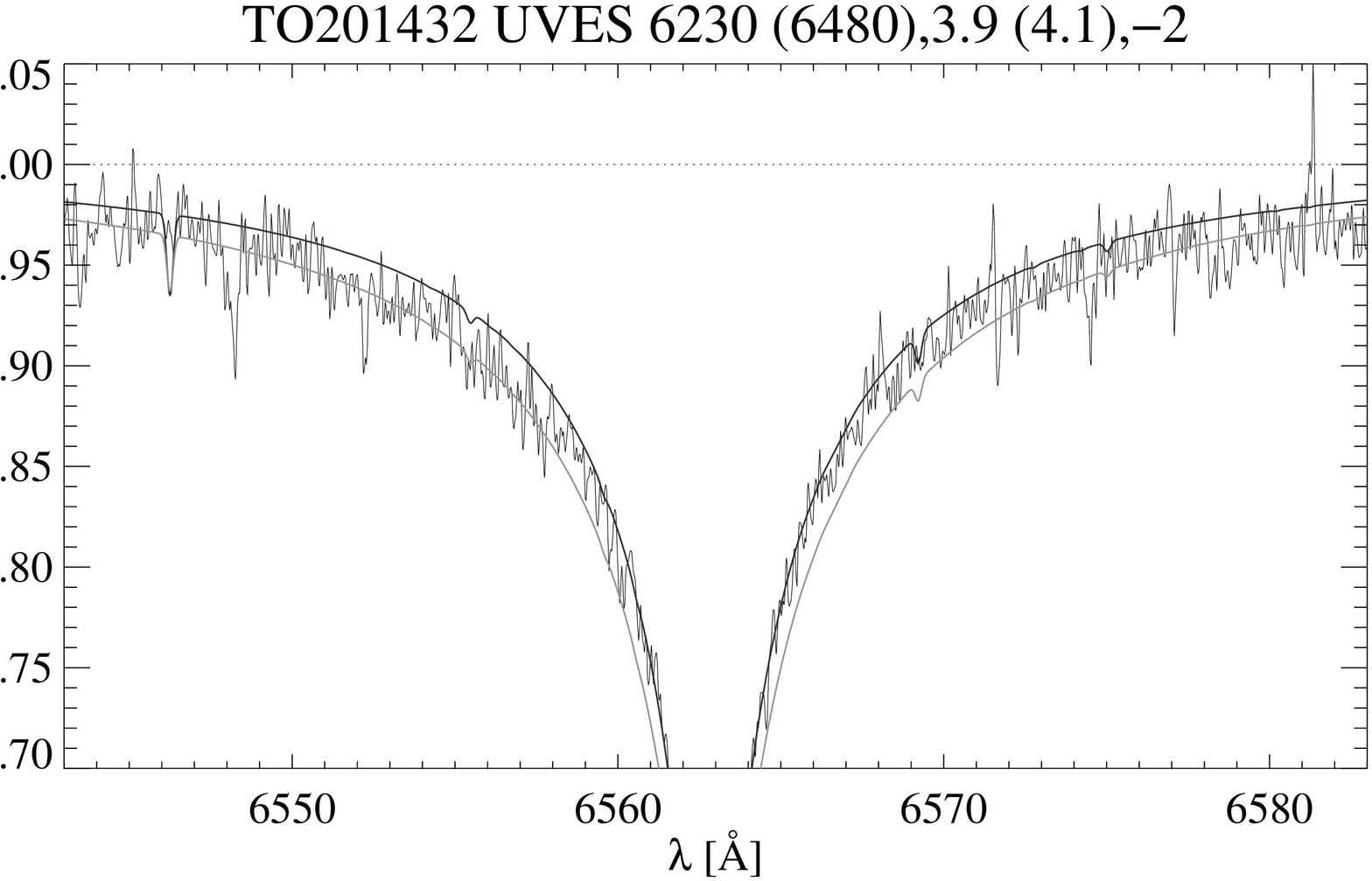}
\end{center}
\vspace*{-1.3cm}
\caption[]{Top panel: Observed and synthetic H$\alpha$ profiles of NGC\,6397/SG793 (black: 5480\,K, grey: $\pm$100\,K). The data was retrieved from the VLT archive (observing dates 2000-06-18 and 20, one spectrum per night), reduced using REDUCE, co-added and rectified using parabolic fits to the continua in adjacent orders. An effective temperature of 5480\,K is indicated, in good agreement with Gratton et al. \cite{gratton}.

Bottom panel: Observed and synthetic H$\alpha$ profiles of NGC\,6397/TO201432 (black: 6230\,K, grey: 6480\,K). The data was retrieved from the VLT archive (observing dates 2000-06-18 and 22, two spectra per night), reduced using REDUCE, co-added and rectified using parabola. An effective temperature of 6480\,K (best estimate of \cite{gratton}) is clearly too high, 6230\,K is indicated instead. This lower effective temperature reduces $\uDkorn\,T_{\rm eff}$(TO\,$-$\,SG) to 750\,K (to be compared with 1000\,K found by \cite{gratton}).}
\label{sgvsto}
\end{figure}

Gratton et al. \cite{gratton} (VLT Large Program 165.L-0.263) pioneered high-resolution, high-S/N observations of turnoff (TO) stars in nearby GCs. In comparing iron abundances in TO and cool subgiant (SG) stars in NGC\,6397 (and NGC\,6752), they found excellent agreement, leaving little room for gravitational settling. Here, we re-examine this claim based on a re-analysis of the archival data using the sophisticated reduction package REDUCE of Piskunov \& Valenti \cite{piskunov}.

Observing TO stars in globular clusters is still a challenging task, even when one has access to 10m-class telescopes and efficient spectrographs like UVES. Gratton et al.~\cite{gratton} use Balmer-profile temperatures as a reddening-free temperature indicator. However, Balmer-profile analyses are challenging, both observationally (Korn \cite{korn}) and theoretically (Barklem et al.~\cite{BPO}). Fig.~\ref{pipeline} shows some of the data used by Gratton et al.~(kindly made available to us by Eugenio Carretta). It is perilous to derive Balmer-profile temperatures from such spectra.

Below, we present an H$\alpha$ re-analysis of the highest S/N targets: the SG\,793 (S/N\,$\approx$\,105) and the TO star 201432 (S/N\,$\approx$\,97, both Fig.~\ref{sgvsto}). We include the halo subgiant HD\,140283 (S/N\,$\approx$\,145, Fig.~\ref{hd140283}), observed in the same LP. With log\,$g$ taken from a 13\,Gyr isochrone, it is the {\em relative} $T_{\rm eff}$ difference ($\uDkorn\,T_{\rm eff}${\small (TO\,$-$\,SG)}) which determines the extent of gravitational settling inferred.

It is particularly interesting to see that we derive a {\em higher\/} effective temperature for the reference field star HD\,140283 (5650\,K, while \cite{gratton} give 5560\,K), but a markedly {\em lower\/} effective temperature for the cluster turnoff star (6230\,K vs. 6480\,K). Seemingly, the subgiant is sufficiently cool (5480\,K) such that the (weak) intrinsic profile of H$\alpha$ could even be recovered from the UVES pipeline spectrum. Observational problems dominate and drastically limit the temperature determination and its reliability.

These results question the effective temperatures assigned to the cluster turnoff stars by Gratton et al. \cite{gratton}. Thus, their conclusions concerning gravitational settling might also be subject to systematic effects. From this work, gravitational settling of iron of up to 0.1\,dex at [Fe/H]\,$\approx$\,$-$2 seems possible.

Clearly, better observations (preferably with a fibre-fed spectrograph) are needed to actually constrain the extent of gravitational settling. Independent temperature indicators (photometry, excitation equilibrium of e.g. Fe\,{\sc i}) should be checked for consistency as well. It is crucial to identify and deal with systematic effects in {\em all\/} aspects of the analysis.

%

\end{document}